# A Systematic Literature Review about Idea Mining: The Use of Machine-driven Analytics to Generate Ideas

Workneh Y. Ayele[1][ 0000-0001-8354-4158] and Gustaf Juell-Skielse [2][ 0000-0002-2922-2286]

[1] Stockholm University, Department of Computer and Systems Sciences, Kista, Sweden

**Abstract.** Idea generation is the core activity of innovation. Digital data sources, which are sources of innovation, such as patents, publications, social media, web-sites, etc., are increasingly growing at unprecedented volume. Manual idea generation is time-consuming and is affected by the subjectivity of the individuals involved. Therefore, the use machine-driven data analytics techniques to analyze data to generate ideas and support idea generation by serving users is useful. The objective of this study is to study state of-the-art machine-driven analytics for idea generation and data sources, hence the result of this study will generally server as a guideline for choosing techniques and data sources. A systematic literature review is conducted to identify relevant scholarly literature from IEEE, Scopus, Web of Science and Google Scholar. We selected a total of 71 articles and analyzed them thematically. The results of this study indicate that idea generation through machine-driven analytics applies text mining, information retrieval (IR), artificial intelligence (AI), deep learning, machine learning, statistical techniques, natural language processing (NLP), NLP-based morphological analysis, network analysis, and bibliometric to support idea generation. The results include a list of techniques and procedures in idea generation through machine-driven idea analytics. Additionally, characterization and heuristics used in idea generation are summarized. For the future, tools designed to generate ideas could be explored.

## 1 Introduction

The elicitation and evaluation of ideas is the core activity of innovation in organizations striving to remain competitive in the global economy. Idea generation, as a creative process, involves human judgment. Also, idea generation is the source of creativity and innovation [1]. The volume of data generated from different sources is increasing at an unprecedented rate. For example, social media, patents, and publications are increasingly growing. Organizations create large volumes of data, yet they usually fail to detect valuable ideas that could lead to innovation [2]. Besides, according to [3], research publications are growing at an unprecedented pace. However, it is hard to generate ideas.

An idea is an abstract concept that is open to interpretations. Thorleuchter et al. referred to ideas as a piece of new and useful text phrase consisting of domain-specific terms from the context of technological language usage rather than unstandardized colloquial language [4]. On the other hand, Liu et al. defined an idea as a pair of problem-solution [5]. In this paper, we refer to "idea" as: "a sentence or text phrase describing novel and useful information through expressing possible solution(s) to current problems."

Idea generation could be done by finding analogies in diverse domains, yet eliciting relevant ideas is difficult [6]. Furthermore, ideas could be generated from diverse sources, as presented in later chapters. Idea mining, which was proposed by [4] in 2010, uses query text and textual data and apply distance-based similarity measures, unsupervised machine learning, to elicit ideas. However, several techniques in literature used other techniques and heuristics; refer to Chapter 2, Related Research and Chapter 4, Results and Analysis. Hence, we refer to these techniques as machine-driven idea generation. Research regarding idea generation supported by computer-aided or machine-driven analytics techniques is fragmented. Besides, to the best of the author's knowledge, machine-driven idea generation probably dates back to around the year 2010 refer to Section 2.2.

In this paper, machine-driven analytics to generate ideas uses not only idea mining, which uses text mining, machine learning, and Information Retrieval (IR), but also bibliometric analysis, NLP-enabled morphological analysis, social network analysis etc. applied on textual data to generate ideas through the use of computers. Machine-driven analytics is helpful in making sense from the increasingly accumulated digital data [7]. Generating ideas from computer-generated patterns obtained from textual data needs analytical reasoning. Endert et al. combined machine learning and visual analytics for sense-making and analytical reasoning [8]. For example, machine-driven analytics supports decision-making, which demands reasoning [9-10]. Similarly, [11] referred to the analysis of textual data using NLP as machine-driven text analytics, and [12] referred to it as machine-driven data analytics.

This paper aims to answer the following research question: *Which type of machine-driven analytics and data types are used to generate ideas?* This paper also explores the state-of-the-art machine-driven idea generation, enlightens the industry and academia to promote innovation, and suggests future research directions. Therefore, a systematic literature review (SLR) over 15 years on selected papers is done. In this paper, a detailed description of mining techniques for idea generation is not presented. Hence, we advise readers from the industry to use this paper as a guideline for choosing techniques and data sources, and readers from academia to explore referred techniques as an inspiration to learning and future research possibilities. This article has six Chapters: Related Research, Methodology, Results and Analysis, Discussions and Limitations, and finally Conclusions.

## 2 Related Research

This paper focuses on ideas generated through machine-driven analytics using digital data sources. Thus, potential digital sources of ideas are patents, scholarly literature [13-14], social media [15], reports, the internet, and documents [13]. Yet, ideas could also be generated using crowdsourcing [2], brainstorming [16-17], crowdfunding [18], etc. A network of experts [19] is also considered a source of ideas.

### 2.1 Machine-driven Analytics for Idea Generation

The research findings of most machine-driven idea generation techniques either overlap with each other or use different names to refer to the same purpose. For example, in computer science, text mining combines data mining, knowledge management, IR, NLP, and machine learning [20]. Besides, machine learning deals with the study of algorithms that learn from experience and improve its performance [21]. Deep learning is part of machine learning that deals with algorithms designed by the inspiration of the human brain's function and structure [22]. Similarly, topic modeling is part of machine learning, which allows the elicitation of hidden topics from textual data [23].

Also, statistics and analytics techniques are also used to generate ideas, in addition to, bibliometric and morphological analysis. Similarly, Social Network Analysis (SNA), which is an interdisciplinary work with concepts established in social theory and application, formal mathematics, statistics, and computing methodology [24]. Likewise, bibliometric is a quantitative assessment of academic publications using statistical methods [25]. Authors use text mining [13,16], data mining [26], AI [17], NLP [13, 27], SNA [28], bibliometric [29], Information Retrieval (IR) [27], deep learning [30], machine learning and experts' feedback [2], machine learning [13, 31], statistical analysis [13], idea mining [4], and topic modeling [32- 33] to generate ideas.

### 2.2 Previous Research

Authors, [34-38] claim that idea mining is introduced by Thorleuchter et al. in 2010 [4]. Thorleuchter et al. used the Euclidean distance-based similarity measuring algorithm to find solutions to queries articulated as problem statements. Ideas are then generated by finding similarities between the articulated problem queries and historical textual data, which is referred to as the distance-based algorithm applied to elicit ideas [4].

However, distance-based algorithms are not the only techniques used to deal with idea generation activities and machine-driven idea generation. For example, [39] used text mining, clustering, and visualization of experts' opinions in computer-supported brainstorming systems for idea generation. Similarly, [31] used time-series analysis and text mining on product attributes of user guides and manuals for generating new ideas in 2005. On the other hand, [40] combined the Analytic Hierarchy Process (AHP), a decision science technique, with data mining models to generate ideas using patents data.

Previous literature reviews mainly focus on using distance-based similarity measuring techniques. For instance, [13] conducted a review of mining ideas from textual data, where they argued that idea mining is a recent field of study and summarized the distance-based algorithms [4]. Similarly, distance-based algorithms are used by [41] and [42]. On the other hand, [43] presented the state-of-the-art creativity supporting systems designed to generate ideas through a literature review. Thus, this article focuses on ideas generated through machine-driven analytics to improve products, services, or research-related activities.

## 3 Methodology

To review state of the art and answer the research question, a systematic literature review (SLR) proposed by [44] and [45] is followed. As suggested by [45], the detailed guideline for conducting SLR consists of three phases: planning, conducting, and reporting. The first phase includes identifying needs, commissioning the review, articulating a research question, developing and evaluating a review protocol. The second phase includes research identification, selection of studies, quality assessment, extraction and monitoring of data, and data synthesis. The last phase includes stating dissemination mechanisms, formatting, and evaluating the report. However, commissioning, evaluating the report, and evaluating the review process are optional [45]. Additionally, to ensure that as many relevant articles as possible are selected, the snowballing technique by [46] is followed. The snowballing technique uses forward and backward searching strategies iteratively [46]. Similar to [44], the SLR methodology includes identification of relevant data sources and search strategies. It also includes selection of articles, quality assessment, data extraction, and performing synthesis based on research protocols.

### 3.1 Planning the SLR

**Need for a systematic review.** According to [13], idea mining is an interesting and new field in information retrieval. Yet, it has gained attention in the data and knowledge engineering domain [47]. Also, idea mining has been successfully used in social studies, medical, and technological domains [41]. Besides, other authors claim that idea mining is introduced by [4], see Section 2.2. The preliminary literature review revealed that machine-driven analytics uses other techniques that are not discussed in the definition of idea mining.

**Specifying the research question.** The research question is articulated to address the need: "*Which type of machine-driven analytics and data types are used to generate ideas?*". Hence, the main purpose of this study is to identify machine-driven analytics and digital data sources that are used for idea generation.

**Inclusion and exclusion criteria.** A pre-defined protocol is formulated to reduce the probability of researcher bias, as suggested in [45]. The inclusion and exclusion criteria are:

- The inclusion criteria used were: (1) publication year (2005-2020), (2) publication outlets (journals articles, and conference proceedings), (3) the selection of publications is based on relevance (papers including idea mining, idea generation, computer-assisted methods, and ideation), (4) databases used (papers published in scientific journals, articles and books published in IEEE, Scopus, Web of Science, and Google Scholar).

- The exclusion criteria: (1) exclude articles that do not qualify the inclusion criteria listed above, (2) workshop papers, book reviews, and cover letters were removed.
- The search strings are derived from the research questions.
- The synthesis of the work should be done to answer the research question.

### 3.2 Evaluation Criteria

We used a predefined protocol to reduce reviewer bias. The protocol enabled us to identify relevant articles using inclusion and exclusion criteria. Additionally, we included peer-reviewed journals and conferences. Included papers were also assessed if they are relevant in terms of answering the research question. Therefore, the papers selected were assessed and checked if they contain machine-driven analytics techniques applied to textual datasets to extract ideas.

According to Kitchenham, a systemic error or bias happens when a researcher fails to produce the "true" result or deviates from it. An unbiased result is considered to be internality valid. On the other hand, applicability or generalizability is external validity, referred to as the study's applicability [44]. In this study, our results are based on the findings we obtained from the literature review, and the results are backed up by citations, and hence it is highly unlikely that we make systemic errors.

### 3.3 Conducting the review

According to [45], the search process must include primary studies and the data extraction process provides information needed to answer the research question. Also, the data analysis process enables the researcher to answer the research question [45]. A snowballing approach is used to make sure that all relevant primary research articles are included. Additionally, to ensure that as many relevant papers as possible are included, the query included synonyms and equivalent terms, as suggested by [45]. The synthesis strategy followed is descriptive synthesis, where the research question is answered through a tabular summary, i.e. an illustrative, and explanatory presentation of the results.

We have been doing a preliminary literature study since January 2018 to explore the research demand. We collected articles for the SLR on December 29, 2019, and the inclusion of relevant articles was done until June 2020. The search query used was: *(((idea AND (generate OR generation OR generating OR elicitation OR elicit OR eliciting OR ideation))) AND (("natural language processing" OR nlp OR clustering OR "topic modeling" OR "information retrieval" OR "data analy\*" OR "social network analysis") OR ( "machine learning" OR "statistical learning" OR "supervised learning" OR "unsupervised learning" ) OR "data mining" OR "text mining" OR "idea mining"))).*

### 3.4 Reporting the review

A total of 1343 from IEEE, 1790 from Scopus, and 2077 from Web of Science articles were extracted. After removing duplicates, a total of 3809 articles remained. After applying inclusion and exclusion criteria and snowballing, a total of 71 articles were selected for the review. The result of the SLR is presented in the next Chapter. The main objective of this study is to explore machine-driven analytics techniques applied fully or partly to generate ideas and data sources used. Besides, this study explores major heuristics for idea generation applied in machine-driven techniques. The evaluation of the systematic literature review is conducted in view of the objective of the research.

## 4 Results and Analysis

In this chapter, the result of the SLR is presented. From the SLR, we understood that idea generation using machine-driven analytics is based on underlying assumptions related to the way ideas are characterized. Therefore, in this chapter, data sources, features characterized show the availability of ideas, machine-driven idea mining techniques are presented. The SLR process is evaluated based on the predefined-protocol and relevance.

### 4.1 Data sources

In this section, data sourced for ideas found in selected publications are presented. Seven different types of data sources appeared only once in selected publications, and these data sources are product descriptions, Wikipedia, sensor data, patents and web, web-blog and documents, web-documents and publications, and patents and publications. The rest of the data sources that appeared in at least two publications are illustrated in Figure 1 below.

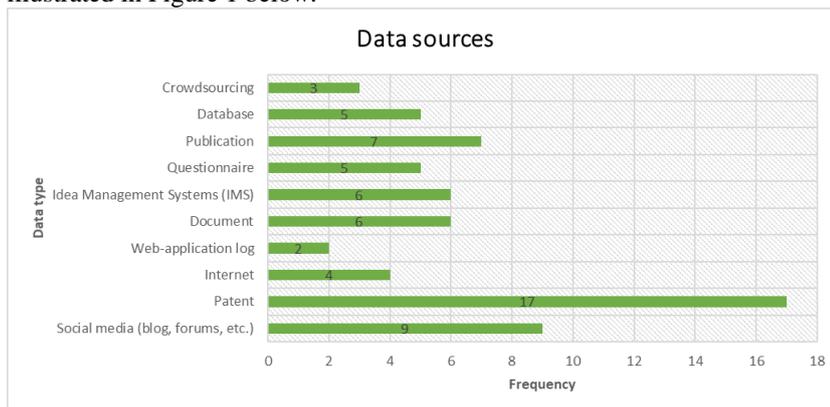

**Fig. 1.** Data sources that appeared in at least two publications where the labels on the bars represent frequency.

## 4.2 Characterization of Textual Data for Idea Extraction

The data sources used for idea extraction include unstructured textual data such as blog posts, social media posts, research publications, etc. which are available in a human-readable format. The characterization of features for identifying ideas are underpinned by assumptions made regarding idea extraction. These assumptions indicate that ideas hidden within textual datasets could be extracted by examining and analyzing the nature of terms or words they contain. For example, textual data extracted from social media was used by [48] to generate creative ideas through the use of suggestive terms, such as "I think", "the solution is", "I suggest", etc. [48].

**Phrases**. Ideas are expressed in phrases, that is, terms containing N-grams (N number of terms). Hence, the elicitation of ideas could employ Part-Of-Speech (PoS) tagging to extract N-grams, phrases with N-words, then apply a technique to assist experts identify new and useful ideas, e.g. [49] used N-gram and Bibliometric, [50] used N-gram to extract concepts and established semantic relationship for idea generation in morphological analysis, a technique discussed below.

**Problem-solution pairs.** The assumption that problems and solutions co-exist demands for techniques to exploit these patterns using techniques such as similarity-based clustering, e.g. distance-based similarity measures, to elicit solutions to problems. For example, phrases (N-grams) describing problems and solutions are used to find inspirational ideas [5]. Suggestive terms, as described above, can be used to elicit association with other terms by applying association mining to discover problem-solution patterns [48].

**Analogy-based (inspiration) idea generation.** Analogical reasoning enables idea elicitation where ideas are generated using inspirations as analogies or solutions designed to solve similar problems. To identify inspirational stimuli during the idea generation process [51] applied Latent Semantic Analysis (LSA) for identifying the extent of semantic similarity between a database containing design examples and current work being done by participants using cosine similarity and participants rating. [52] applied (text mining) cosine similarity between F-term, technical attributes to classify patent documents, and patent document to select a target technology for suggesting new technology ideas. Similarly, [53] indicated that out-of-domain knowledge, where they referred to it as distant analogies, stimulates idea generation. Also, pioneers benefit from using analogies to generate innovative ideas through searching for analogies, IR [53]. Hence, [54] applied analogical designing, clustering of patents having similar design cases, to prompt new product design idea generation [54].

**Trend based or time-series analysis-based idea generation.** Topic evolutions are generated using dynamic topic modeling using scholarly articles, and succeeding statistical time-series analysis and visualizations enable the elicitation of research and innovation ideas using [33]. Similarly, [31] used text mining and time-series analysis on product attributes of user guides and manuals for generating new ideas.

### 4.3 Idea mining techniques

Through the use of the SLR, major machine-driven analytics for idea generation techniques are identified as illustrated in Table 1.

**Table 1.** List of machine-driving analytics techniques for idea generation, where **Idea Type** - P: Product, G: General, Pr: Process, M: Marketing, PD: Product Design, T: New Technological Idea, R: Research, C: Concept, S: Service.

| Techniques of Idea Gene | Data Types | Algorithms or Methods | Idea Type | Authors |
|---|---|---|---|---|
| **Computer-driven text analysis for generating ideas** | | | | |
| Social Network Analysis | Social media | Semantic Network Analysis (SNA) is applied to build word co-occurrence network, and [55] indicated that networks communicating agents indicate the existence of novel ideas, the existence of preferences, etc. | G | [55] |
| | Blog | Network analysis using two models, i.e., semantic ideation network and the visual concepts combination model for providing computational creativity through semantics and visualizations. | P | [56] |
| | News websites (discourse data) | Text analytics and semantic networks of concepts are used to generate ideas and facilitate open innovation from technological discourse posted on websites. | G | [57] |
| | IMS | Semantic networks analysis of (words) and visualization of co-occurrence are used to generate ideas. | G | [58] |
| | Internet log file | Lexical Link Analysis (LLA), which is unsupervised machine learning that generates semantic networks, is applied on an internet gaming log file to generate ideas. | Game | [59] |
| | Social media data | Named entity recognition and PoS are used to preprocess the data, and the social media idea ontology graph is generated to inspire idea generation. | P/M | [60] |
| | Patent | Co-word and network analysis of keywords and visualization using  Multi-Dimensional Scaling (MDS). | P/M | [61] |
| Bibliometric Analysis | Publications from  Scopus | Co-word and co-citation analysis using N-grams extracted through PoS is used to generate ideas. | G | [49] |
| | Patent | Patent citation network analysis using keywords based on outlier detection through the assessment of centrality measures. | G | [62] |
| Information Retrieval (IR) | Internet (Web) and Patent | A function-based IR, which uses k-means clustering and classification on patent data organized using TF-IDF features, is used to suggest solutions to problems using a problem-solution heuristic to inspire idea generation. | PD | [63] |
| | Patent | Problem-solution heuristics is used for searching mechanisms or solutions to achieve the purpose. | P | [54] |
| | Patent | Analogy based IR method that applies deep learning and crowdsourcing to look for mechanisms for purposes. | P | [30] |
| | database (product design) | Using Big data and applying Latent Semantic Analysis (LSA) for idea generation. | PD | [64] |
| | Social media data crowdsourcing | The use of customized databases and WordNet is limited as both have limitations in number and variety of data. Hence authors suggested the use of NLP and data mining to support preprocessing crowd-knowledge. | PD | [26] |

**Supervised machine learning**

| | | | | |
|---|---|---|---|---|
| **Classification** | Document (technological) | Using a database containing analogies and a text classification algorithm to inspire idea generation | P | [65] |
| | IMS-LEGO | SVM, nearest neighbor, decision tree, neural networks – applied on labeled data to demonstrate identification of ideas | PD | [66] |
| | IMS –LEGO & Beer | SVM and Partial Least Squares classifiers | P/Pr/M | [67] |
| | IMS | Sentiment analysis, term features extracted sung TF-IDF. Sentiment analysis using SentiWordNet for extracting ideas based on probability of adoptability | P/M | [68] |
| **k-NN** | IMS –DELL (Dell IdeaStorm) | Outlier detection using k-NN applied on a document term matrix represented as TF-IDF using cosine based distance measure | P/S | [68] |
| **Regression and time series analysis** | IMS | NLP and Logistic regression | P | [70] |
| | Document (product manuals) | Vector auto-regression (VAR)  model | PD | [31] |
| | Publication | Using Dynamic Topic Modeling (DTM) evolution of topics is identified then time series, correlation …… prediction on topic evolution generated using DTM | T/R | [33] |

**Unsupervised machine learning**

| | | | | |
|---|---|---|---|---|
| **Clustering** | Patent and web report | VSM, Cosine similarity, ORCLUS[1] | P | [71] |
| | Crowdsourcing | EM-SVD and HDBSCAN algorithms and human evaluators (analogy) | Ct | [72] |
| | Web document and publications | Concept clustering using similarity measures between to find concept association | PD | [73] |
| | Patent | Clustering using cosine-similarity applied on F-term (patent classification information) and patent documents | T | [52] |
| | Patent and publication | To identify  gaps and analyze the correspondence between technology and science **authors** used ORCUS clustering | T | [74] |

---

[1] Arbitrary Oriented Cluster Generations

| | | | | |
|---|---|---|---|---|
| **Association Mining** | Social media forums | Apriori association mining algorithm | P/S | [48] |
| | Database (data-ware-house) | Fuzzy ARM | P | [75] |
| | Question-naire | Apriori association mining algorithm | P/M | [76] |
| | Database (customer and transac-tion data) | Apriori association mining algorithm | P/M | [77] |
| | Patent | Apriori association mining algorithm | P/M | [78] |
| **Dimension Reduction and similar** | Patent | Principal Component Analysis(PCA) | P/T | [18] |
| | Patent | Generative Topographic Mapping is better than PCA in terms of visualization according to [73] | T | [79] |
| | Patent | NLP-LSA and outlier detection | P | [80] |
| | Patent | {27} [78] (Yoon & Kim, 2012) applied WordNet on patent data to identify Subject-action-object (SOA), example "(S) mobile (O) has (A) battery", semantic analysis based on semantic similarity, and multidimensional scaling visualization to detect outliers whereby new technological ideas could be elicited. | T | [81] |
| **Topic modeling Co-(occurrence analysis)** | IMS using online crowdsourc-ing | Co-occurrence analysis and visualization | P/S | [82] |
| | database (product de-sign) | Using Big data and applying LSA | PD | [64] |
| | Social media | LDA is used to identify latent product topics from customer-generated social media data and then applying sentiment analysis to measure satisfaction level. | P/M | [83] |
| | Publication | DTM to generate evolution of topics and visualization of trends | T/RS | [33] |

## Combined techniques

| | | | | |
|---|---|---|---|---|
| **Recommendation sys-tems** | Publications | PoS tagging is for detecting noun-phrases within titles and abstracts, through which problem-solution pairs and adding inverse-document-frequency to make it triplets for collaborative filtering algorithms where problem phrases are treated as users and solution-phrases are treated as recommended items | G | [5] |
| | Patent | LDA and collaborative filtering to recommend for exploring opportunities, and visualization on recommended products to identify new application products | P | [84] |
| **Mis-cellane-ous** | Question-naire | Apriori and Clustering using K-medoid and RPglobal | S | [85] |
| | Question-naire | ARM (Apriori) and Clustering (K-means) | P/M | [86] |

| Social media and computer-generated data | ARM and distance-based Clustering | S | [87] |
|---|---|---|---|
| Patent | Latent Dirichlet Allocation (LDA) and ARM | P | [88] |
| Questionnaire and interview | Apriori, K-means clustering, hill-climbing and density-based DBSCAN partitioning | P | [89] |
| Database of examples files | LSA and semantic similarity measure (cosine similarity) to find similar solutions for a given problem (analogy-based) | PD | [51] |
| -Questionnaire and survey | Apriori (ARM) and C5.0 (decision tree) | P | [90] |
| Websites (crawled) | ARM and Decision tree | P | [91] |
| Questionnaire and interview | Apriori (ARM) and C5.0 (decision tree) | P/M | [92] |
| web sources about renewable energy | The idea mining technique proposed by [4] was used to generate ideas, and clustering using latent semantic indexing was used to create semantic clusters, and finally, to classify concepts, Jaccard's similarity was used to identify interdisciplinary idea. | G | [93] |
| Social media | Idea generation through the combination of topic modeling, LDA, and sentiment analysis to measure satisfaction. | P/M | [83] |

### 4.4 Summary of Identified Idea Generation Techniques Using Idea Mining

As discussed in Section 2.2, idea mining was introduced by [4], where distance-based similarity measures were used to identify ideas. However, in this study, we have also identified that visualization, dimension reduction, NLP-driven morphological analysis, bibliometric, scientometric, and network analysis are used for idea generation. It is also found that opinion mining, idea mining, and topic signal mining are applicable for idea generation [94].

**Visualization and dimension reduction.** [80] developed a visualization tool using the NLP technique, LSA, to detect outliers for product idea generation using patent data. Also, visualization of PCA is used to detect technology gaps [18]. Unstructured data using circular layout and Force-Directed graph (CiFDAL) algorithms are applied to generate scenario graphs to generate ideas [95]. Visualization of co-occurrence graphs can be used to generate ideas [82], for example, visualization of a patent citation network of keywords to generate ideas by combining expert knowledge [62]. The use of NLP in combination with WordNet-based semantic similarity measure to extract terms that are subject-action-objects for outlier detection using a keyword relationship-based visualization algorithm called multidimensional scaling (MDS) [81]. Outlier detection using data collected from Dell IdeaStorm and My Starbucks Idea,

IMS, were used clustering to identify outliers using k-NN and TF-IDF based features for generating product and service ideas [68]. [81] applied WordNet on patent data to identify subject-action-object (SAO), for example "(S) mobile (A) has (O) battery," semantic analysis based on semantic similarity, and multidimensional scaling visualization to detect outliers whereby new technological ideas could be elicited.

**Morphological analysis** is a method employed to generate ideas through breaking down a system into dimensions for the purpose of comprehensively elaborating the ins and outs of the system [96]. According to [97], the success of morphological analysis critically depends on its dimensions. However, the process of building morphological structures and values is significantly affected by subjectivity and bias, hence [97] proposed the use of WordNet, a dictionary of a lexical database consisting of a hierarchical network of words, for morphology building to establish dimensions and values. The morphological building process uses a hierarchy of words, WordNet, consisting of meronym (a semantic relation showing A is a member or part of B, if A denotes the whole part of B, e.g., faces are used to mean people), holonym[2] (a semantic relation showing a term and its relation representing it as part of whole, e.g., a finger is part of a hand), hyponym (a term which is more specific than a general term applicable to it, e.g., screwdriver is a hyponym of toolset), and hypernym (hypernym is the opposite of hyponym, e.g., color is a hypernym of green) as defined in Oxford University Press[3] for dimension and value construction. Similarly, [98] demonstrated that it is possible to build a morphological matrix using Word2Vec clustering analysis and F-term for generating ideas using patent data.

Text mining is used to support the building of morphological analysis, text-driven morphological analysis, c.f. [15, 96, 99]. Geum et al. used sensor data to generate ideas [97]. The limitation of morphological analysis is that it is difficult for experts to evaluate a multitude of generated product, concept, or service ideas, therefore, [96] applied conjoint analysis, a statistical method for measuring customers' feedback among product/service idea attributes. A combination of morphological, conjoint, and citation analyses can also be used to generate new technological ideas from patents [100]. Finally, [50] used morphological analysis using Wikipedia for generating ideas using basic text analysis techniques.

**Bibliometric** is a quantitative assessment of academic publications using statistical methods [35]. It is possible to use publications as a data source for bibliometric analysis. Also, bibliometric combined with link mining and text mining, it is possible to elicit research ideas [23]. Co-citation analysis of academic publications can be done to elicit ideas [101]. Co-word and co-citation analysis using N-grams extracted through PoS (Text mining, Stanford PartOfSpech) tagger to extract N-grams by using abstracts of research publications and removing noisy words, selecting N-grams having potential ideas, using Scopus database to check if candidate ideas are useful and innovative ideas [49].

---

[2] https://en.wiktionary.org/wiki/holonym
[3] https://www.lexico.com/definition

**The use of enhanced IR for searching ideas.** To aid mechanical design idea generation, [64] applied AI and big data analytics to train mechanical design using LSA for predicting the relevance of retrieved design elements for an articulated query thereby retrieving previous designs to the user. Liu et al., created a local databased through crawling the internet to extract international patent information, functional information, and patent documents through which they created function-based patent IR [97]. The function-based IR tool uses a semi-supervised machine learning to map function problem space with design problem space. They showed that function-based knowledge IR systems outperform standard IRs [63]. Similarly, [54] by applying NLP on patent data created an analogical source of knowledge through the development of VSM for supporting peoples creatively to generate ideas.

Patent based IR systems also support idea generation through the construction of a VSM with problem-solution space representation of patent data through crowdsourcing and deep learning [30]. The evaluation of analogy-based retrieval systems yields better performance in terms of precision and recall than traditional IR systems [30]. [65] demonstrated that a database created to store analogical information supports users in retrieving relevant analogical information that facilitates novel idea generation. [26] for product design – social media data was used, and the use of customized database and WordNet is limited as both have limitations in number and variety of data. Hence authors suggested the use of NLP and Data Mining to support the preprocessing of crowd-knowledge.

## 5        Discussions and Limitations

The purpose of this paper was to explore idea generation techniques. The SLR indicates that there are many techniques to support idea generation using machine-driven data analytics. Ideas are hidden within textual datasets, and these digitally available data sources could be social media data, websites, different types of documents containing technical information, patents, scholarly articles, application logs, sensor data, a compilation of idea generation targeted questionnaire and interview data, databases of customer information and transactions, and documents collected during crowdsourcing events. Similarly, IMS are also sources of ideas, for example Dell IdeaStorm, My Starbucks Idea [68], LEGO, and an IMS used by a US-based Brewery [67]. The most widely used digital source of ideas in order of frequency are patents, social media, scholarly articles, IMS, questionnaires, databases, and the Internet in general. Also, the most widely used techniques of mining textual data for idea generation are – ARM using Apriori, clustering, a combination of ARM and clustering, co-occurrence analysis, dimension reduction, network analysis, and regression.

Today, machine-driven data analytics for generating ideas is demanded because it takes time to manually process large volumes of textual data. Computers need to be programmed to act like humans in order to process large volumes of data for simplifying and expediting laborious human tasks. The result of this study enabled us to define, machine-driven analytics for idea generation as "*the use of digital data sources, such* *as textual, to generate innovative ideas using text mining, machine learning, NLP, statistical analysis, IR, deep learning, topic modeling, social network analysis, bibliometric, NLP-based morphological analysis, and visual analytics*". The most commonly used features and assumptions underpinning idea generation through machine-driven approaches are the types of words that indicate the availability of ideas, the association between words or phrases, the existence of analogies, or solutions for a given set of problems or queries, and the existence of trends and patterns. According to [48], suggestive terms such as – "I think", "The result", "I assume", etc. and their association with other terms existing in the datasets could be used to generate ideas. We also learned that analogy based searching using datasets or IR based databases expedites idea generation. Noun phrases could spur the generation of ideas. For example, text-phrases in bibliometric [49], text-phrases in morphological analysis [50], and text-phrases in problem-solution elicitation [5] are used for generating ideas. Time-series analysis is valuable for the elicitation of trends, temporal patterns, insights, and foresight, which in turn spur innovation. However, more works need to be done regarding the trend-driven idea generation. The result of this study could be used to support idea generation by serving as a guideline.

This study has limitations which needs to be addressed in future studies. The synthesis of selected papers is mainly focused on identification of techniques, data sources, characterizations, and heuristics. Hence, it overlooks a deeper analysis of techniques with respect to application areas. Similarly, as the purpose of this paper is mainly focused to identify techniques and data sources, a deeper analysis of aspects such as evaluation of employed techniques, heuristics involved is overlooked. For the future, it is recommended to conduct an SLR with a deeper synthesis.

## 6    Conclusions

The study of machine-driven analytics for idea generation is fragmented. For example, some consider idea generation to involve distance-based similarity to measure the semantic closeness between problem queries and historical data, while others consider…. We identified more than twenty machine-driven analytics techniques in the literature. Besides, we identified the characterization or features that are used for idea generation, data sources, and underlying heuristics.

Practitioners can use this study as a source of idea generation toolbox and relevant data sources. Thus, this study could summarize techniques and heuristics for idea generation for enthusiastic practitioners. It could also serve as a basis for future work in academia. Ideas are sources of innovation, and hence the result of this study is a valuable contribution to the industry, startups, accelerators, incubators, government agencies, etc.